\documentclass[showpacs,amssymb,aps,twocolumn]{revtex4}
\usepackage{amsmath}
\usepackage{graphicx}
\usepackage{epsf}
\usepackage{graphics}
\usepackage[latin1]{inputenc}
\newcommand{\be}{\begin{equation}}
\newcommand{\ee}{\end{equation}}
\newcommand{\bea}{\begin{eqnarray}}
\newcommand{\eea}{\end{eqnarray}}

\newcommand{\vp}{\vspace{0.3cm}}

\newcommand{\undertilde}[1]{\underset{\widetilde{}}{#1}}

\begin{document}
  
\title{General Covariant Gauge Fixing for Massless Spin-Two Fields}

\author{F. T. Brandt$^{a}$\footnote{fbrandt@usp.br},
J. Frenkel $^{a}$\footnote{jfrenkel@fma.if.usp.br}
and D. G. C. McKeon$^{b}$\footnote{dgmckeo2@uwo.ca} }
\affiliation{$^a$ Instituto de F\'{\i}sica,
Universidade de S\~ao Paulo,
S\~ao Paulo, SP 05315-970, Brazil}
\affiliation{$^b$ Department of Applied Mathematics, The University of
Western Ontario, London, ON  N6A 5B7, Canada}

\vp 

\begin{abstract}
The most general covariant gauge fixing Lagrangian is considered for a
spin-two gauge theory in the context of the Faddeev-Popov
procedure. In general, five parameters characterize this gauge
fixing. Certain limiting values for these parameters give rise to a
spin-two propagator that is either traceless or transverse, but for no
values of these parameters is this propagator simultaneously traceless
and transverse. Having a traceless-transverse (TT)
propagator ensures that only the physical degrees of freedom associated
with the tensor field propagate, and hence it is analogous to the Landau
gauge in electrodynamics. To obtain such a traceless-transverse propagator,
a gauge fixing Lagrangian which is not quadratic must be employed;
this sort of gauge fixing Lagrangian is not encountered in the usual
Faddeev-Popov procedure. It is shown that when this non-quadratic
gauge fixing Lagrangian is used, two Fermionic and one Bosonic ghost
arise. As a simple application we discuss the energy-momentum tensor of
the gravitational field at finite temperature.
\end{abstract}


\maketitle

\section{Introduction}

The quantum mechanical path integral provides a useful way of
quantizing gauge field theories as the contributions of superfluous
gauge degrees of freedom to physical process can be cancelled by the
contribution of ``ghost'' fields without breaking general covariance
\cite{Feynman:1963ax,DeWitt:1967yk,Faddeev:1967fc,Mandelstam:1968hz}.
A degree of arbitrariness in this procedure occurs, as one must at the
outset choose a particular ``gauge fixing'' Lagrangian, although
physical quantities are necessarily independent of this choice.

A spin one field $A_\mu$, even when it is not a gauge field (i.e. it
is a ``Proca field''), satisfies the transversality condition 
\be\label{e1}
\partial\cdot A = 0
\ee 
so that it has only the three degrees of freedom normally associated
with spin-one. In order to restrict the propagating degrees of freedom to those that are
physical, it is often convenient that the propagator for a
spin-one gauge field $D_{\mu\nu}(k)$ is also taken to be transverse so
that
\be\label{e2}
k^\mu D_{\mu\nu}(k) = 0.
\ee
This condition is satisfied in the so-called ``Landau gauge'' in which
the quadratic gauge fixing Lagrangian
\be\label{e3}
{\cal L}_{\mbox{gf}} = -\frac{1}{2 \alpha}\left(\partial \cdot A\right)^2 
\ee
is used and the limit $\alpha\rightarrow 0$ is taken.

A spin-two field is associated with a symmetric tensor field $h_{\mu\nu}$;
in order for it to have five independent degrees of freedom it must be
both traceless and transverse
\begin{eqnarray}\label{e4}
\eta^{\mu\nu} h_{\mu\nu} &=&0 \;\;\;\; (\eta^{\mu\nu}\equiv{\mbox{diag}}(+\,-\,-\,-))\\
\label{e5}
\partial^\mu h_{\mu\nu} &=&0 .
\end{eqnarray}
When this field $h_{\mu\nu}$ becomes a gauge field and self-coupled to
its own energy-momentum tensor, it is identified with the graviton 
\cite{tHooft:2002xp}. It is often convenient to use a ``TT
propagator'' $D^{\rm TT}_{\mu\nu,\lambda\sigma}(k)$ which satisfies
\begin{eqnarray}\label{e6}
\eta^{\mu\nu} D^{\rm TT}_{\mu\nu,\,\lambda\sigma}(k) &=&0 \\
\label{e7}
k^{\mu} D^{\rm TT}_{\mu\nu,\,\lambda\sigma}(k) &=&0 .
\end{eqnarray}
In  refs. \cite{Gribosky:1988yk,Rebhan:1990yr,Brandt:2007zz}, 
such a gauge proves to be quite useful when dealing with
the thermal properties of the gravitational field.
In this paper we explain how such a propagator arises when using the
path integral quantization.

We begin by examining the Faddeev-Popov procedure for quantizing gauge
theories using a more transparent matrix analogue for illustrative
purpose. We then apply this procedure to a spin-two gauge field, using
the most general covariant quadratic gauge fixing Lagrangian
possible. We show how a traceless propagator (satisfying \eqref{e6})
and a transverse propagator (satisfying \eqref{e7}) can occur, while
it is impossible to obtain a propagator that satisfies both
Eqs. \eqref{e6} and \eqref{e7}.

Next, the Faddeev-Popov procedure is generalized to accommodate a
non-quadratic gauge fixing Lagrangian. It is shown how such a
Lagrangian can be used to give rise to 
$D^{\rm TT}_{\mu\nu,\,\lambda\sigma}$ satisfying 
Eqs. \eqref{e6} and \eqref{e7}. Three ghost fields occur in this
procedure, two Fermionic and one Bosonic. 
In the last section, we calculate the leading temperature corrections
to the energy-momentum tensor and confirm that the result, which has
been previously obtained, is gauge invariant.

\section{The  Faddeev-Popov Procedure and Covariant Gauge Fixing for Spin-Two}

If we consider the standard integral
\be\label{e8}
Z = \int {\rm d}\vec h \;\exp{\left(-\vec h^T \undertilde{M} \vec h \right)} 
=\frac{\pi^{n/2}}{\det^{1/2}\undertilde{M}}
\ee
where $\vec h$ is an $n$-dimensional vector, it is understood that all
eigenvalues of the matrix $\undertilde{M}$ are positive definite.
If there exists a matrix $\undertilde{A}^{(0)}$ such that
\be\label{e9}
\undertilde{M} \undertilde{A}^{(0)} \vec\theta = 0
\ee
for any given vector $\vec\theta$, then $\undertilde{M}$ has vanishing
eigenvalues and Eq. \eqref{e8} is ill defined.  The Faddeev-Popov 
\cite{Faddeev:1967fc,Hooft:1971fh} procedure for ascribing a meaning
to Eq. \eqref{e8} when this problem arises involves first inserting
\be\label{e10}
1 = \int {\rm d}\vec\theta
\delta(\undertilde{F}(\vec h + \undertilde{A}^{(0)} \vec\theta) - \vec p) 
\det(\undertilde{F} \undertilde{A}^{(0)})  
\ee
into Eq. \eqref{e8}, and then making the change of variable
\be\label{e11}
\vec h\rightarrow \vec h - \undertilde{A}^{(0)} \vec\theta
\ee
leaving us with
\be\label{e12}
Z = \int {\rm d}\vec\theta\int {\rm d}\vec h 
\delta(\undertilde{F} \vec h  - \vec p) 
\det(\undertilde{F} \undertilde{A}^{(0)})  
\exp{\left(-\vec h^T \undertilde{M} \vec h \right)} ,
\ee
where we have used Eq. \eqref{e9}. If now a factor of
\be\label{e13}
1 = \pi^{-n/2}\int {\rm d}\vec p\,
{\rm e}^{-\vec p^{T} \undertilde{N} \vec p}{{\det}^{1/2}(\undertilde{N})}
\ee
were inserted into Eq. \eqref{e12}, then we would be left with
\begin{eqnarray}\label{e14}
Z & = & \pi^{-n/2}
\int{\rm d}\vec\theta\int{\rm d}\vec h 
\det(\undertilde{F} \undertilde{A}^{(0)}){{\det}^{1/2}(\undertilde{N})}
\nonumber \\ & &
\exp{\left[-\vec h^T\left(\undertilde{M}+\undertilde{F}^{T}\undertilde{N}\undertilde{F}\right)\vec h\right]}.
\end{eqnarray}
Exponentiating the determinants occurring in Eq. \eqref{e14} using
Grassmann ``ghost''fields leads to
\begin{eqnarray}\label{e15}
Z & = & \pi^{-n/2}
\int{\rm d}\vec\theta\int{\rm d}\vec h \int{\rm d}\vec{\bar c} \int{\rm d}\vec c\int{\rm d}\vec k
\nonumber \\ & & \!\!\!\!\!\!\!\!\!\!\!\!\!\!\!\!\!\!\!\!
\exp{\left[-\vec{\bar c}\undertilde{F}\undertilde{A}^{(0)}\vec c
-\vec k^{T}\undertilde{N}\vec k
-\vec h^T\left(\undertilde{M}+\undertilde{F}^{T}\undertilde{N}\undertilde{F}\right)\vec h\right]}.
\end{eqnarray}
The Faddeev-Popov ghosts are $\vec c$ and $\vec{\bar c}$; $\vec k$ is
a Nielsen-Kallosh ghost \cite{DeWitt:1967yk,Nielsen:1978mp,Kallosh:1978de}. The
``infinity'' occurring in Eq. \eqref{e8} as a result of $\det\undertilde{M}$
vanishing now is parametrized by the integral over the ``gauge function''
$\vec\theta$ which can be absorbed into a normalization factor.

For a spin-two gauge field, we take the second order term in the
Einstein-Hilbert action to be the classical action so that
\be\label{e16}
S = - \int{\rm d}^d x\left(h^{\lambda\sigma} M_{\lambda\sigma,\mu\nu} h^{\mu\nu}\right)
\ee
where in momentum space
\begin{eqnarray}\label{e17}
M_{\lambda\sigma,\,\mu\nu} & = &\frac{k^2}{2}\left[
\frac{1}{2}\left(\eta_{\mu\lambda}\eta_{\nu\sigma}  + \eta_{\nu\lambda}\eta_{\mu\sigma} \right)
-\eta_{\mu\nu}\eta_{\lambda\sigma}\right]
\nonumber \\
&-&\frac{1}{4}\left[
k_\mu k_\lambda \eta_{\nu\sigma}+k_\nu k_\lambda \eta_{\mu\sigma}
+k_\mu k_\sigma \eta_{\nu\lambda}+k_\nu k_\sigma \eta_{\mu\lambda}\right]
\nonumber \\
&+&\frac{1}{2}\left[
k_\mu k_\nu \eta_{\lambda\sigma}+k_\lambda k_\sigma \eta_{\mu\nu}
\right]. 
\end{eqnarray}
This is invariant under the gauge transformation
\be\label{e18}
\delta h_{\mu\nu} = \partial_\mu \theta_\nu  + \partial_\nu \theta_\mu \equiv
{A}^{(0)}_{\mu\nu\,  \lambda} \theta^\lambda, 
\ee
where
${A}^{(0)}_{\mu\nu\,  \lambda} = \eta_{\nu\lambda}\partial_\mu+\eta_{\mu\lambda}\partial_\nu$. 
The most general covariant ``gauge fixing'' condition is
\begin{eqnarray}\label{e19}
\undertilde{F}\vec h & = & F_\alpha^{\;\;\lambda\sigma}\, h_{\lambda\sigma}
\nonumber\\ & = &
\left[
\frac{1}{\alpha} k_\alpha \eta^{\lambda\sigma} + \frac{1}{\beta}
\left(k^\lambda \delta_\alpha^\sigma + k^\sigma \delta_\alpha^\lambda\right)
\right.\nonumber \\ &+&\left.
\frac{1}{\gamma}\frac{k_\alpha k^\lambda k^\sigma}{k^2}
\right] h_{\lambda\sigma}
\end{eqnarray}
so that the ``gauge fixing'' Lagrangian is
\be\label{e20}
{\cal L}_{\mbox{gf}} =  - h_{\lambda\sigma}F_\alpha^{\;\;\lambda\sigma} 
N^{\alpha\beta} F_\beta^{\;\;\mu\nu} h_{\mu\nu}
\ee
where the ``Nielsen-Kallosh'' factor is
\be\label{e21}
N^{\alpha\beta} = \xi \eta^{\alpha\beta} + \zeta \frac{k^\alpha k^\beta}{k^2}. 
\ee
In the special case when $\gamma\rightarrow\infty$, $\xi=1$ and $\zeta=0$ this general class 
of gauges reduces to the one considered in \cite{Nishino:1977pw} where
the spin-two propagator was considered in various limits of the gauge parameters.

Upon introducing
\begin{subequations}\label{e22}
\begin{eqnarray}
T^1_{\lambda\sigma,\,\mu\nu} & = & \eta_{\mu\lambda}\eta_{\nu\sigma}+\eta_{\nu\lambda}\eta_{\mu\sigma}
\\
T^2_{\lambda\sigma,\,\mu\nu} & = & \eta_{\mu\nu}\eta_{\lambda\sigma}
\\
T^3_{\lambda\sigma,\,\mu\nu} & = & \frac{1}{k^2}\left(k_\mu k_\lambda \eta_{\nu\sigma} +
                              k_\mu k_\sigma \eta_{\nu\lambda}\right) + (\mu\leftrightarrow\nu) 
\\
T^4_{\lambda\sigma,\,\mu\nu} & = & \frac{1}{k^2}\left(
k_\mu k_\nu \eta_{\lambda\sigma} + k_\lambda k_\sigma \eta_{\mu\nu} \right)
\\
T^5_{\lambda\sigma,\,\mu\nu} & = & \frac{1}{k^4}\left(k_\mu k_\nu k_\lambda k_\sigma\right)
\end{eqnarray}
\end{subequations}
then Eq. \eqref{e20} becomes
\begin{eqnarray}\label{e27}
L_{\mbox{gf}} & = & -h^{\lambda\sigma}\left\{
\frac{\xi+\zeta}{\alpha^2} T^2_{\lambda\sigma,\,\mu\nu}
+\frac{\xi}{\beta^2} T^3_{\lambda\sigma,\,\mu\nu}
\right . \nonumber \\ & + & \left.
\frac{\xi+\zeta}{\alpha}\left(\frac{2}{\beta}+\frac{1}{\gamma}\right)
T^4_{\lambda\sigma,\,\mu\nu}
\right . \nonumber \\ & + & \left.
\left[\frac{\xi+\zeta}{\gamma}\left(\frac{4}{\beta}+\frac{1}{\gamma}\right)
+\frac{4\zeta}{\beta^2}\right]{T^5_{\lambda\sigma,\,\mu\nu}}
\right\} {k^2} h^{\mu\nu}. 
\end{eqnarray}
The propagator for the spin-two field with this gauge fixing
Lagrangian is given by $D_{\lambda\sigma,\, \alpha\beta}$
\begin{eqnarray}\label{e28}
D^{\lambda\sigma, \, \alpha\beta} \left(M_{\alpha\beta,\,\mu\nu} + 
F_{\rho,\alpha\beta} N^{\rho\delta} F_{\delta,\mu\nu}\right)
\nonumber \\
=\frac{1}{2}\left(\delta_\mu^\lambda\delta_\nu^\sigma+
                  \delta_\nu^\lambda\delta_\mu^\sigma\right)
\equiv \bar\Delta^{\lambda\sigma}_{\mu\nu}.
\end{eqnarray}
Explicit calculation leads in $d$ dimensions to
\begin{eqnarray}\label{e29}
D_{\mu\nu, \, \lambda\sigma}(k) = \frac{1}{k^2}\sum_{i=1}^5 {\bf C}^i T^i_{\mu\nu,\, \lambda\sigma}, 
\end{eqnarray}
where 
\begin{subequations}\label{e30}
\begin{eqnarray}
{\bf C}^1 & = & 1 
\\
{\bf C}^2 & = & -\frac{2}{d-2}  
\\
{\bf C}^3 & = & \left(\frac{\beta^2}{4\xi}-1\right) 
\\
\label{e33}{\bf C}^4 & = & \frac{2}{d-2}
\left[1+\frac{\beta\gamma}{\alpha(\beta+\gamma)+\gamma(\alpha+\beta)}\right]
\\
{\bf C}^5 & = & -\frac{\beta^2}{\xi} 
+ \frac{1}{\xi+\zeta}\frac{(\alpha \beta \gamma)^2}{[\alpha(\beta+\gamma)+\gamma(\alpha+\beta)]^2}
\nonumber \\ 
&+& \frac{2}{d-2}\frac{(d-3)\alpha(\beta+2\gamma)-2\beta\gamma}{\alpha(\beta+\gamma)+\gamma(\alpha+\beta)}.
\end{eqnarray}
\end{subequations}
For comparison, we note that the analogous propagator for a spin-one
gauge field when using the gauge fixing Lagrangian 
$\bar{\cal L}_{\mbox{gf}}=-\frac{1}{2\alpha}(\partial\cdot A)^2$ is
\begin{eqnarray}\label{eq35}
\left[-k^2\eta^{\mu\nu}+\left(1-\frac{1}{\alpha}\right)k^\mu k^\nu\right]^{-1}
\nonumber \\
=\left(-\frac{\eta_{\mu\nu}}{k^2}+(1-\alpha)\frac{k_\mu k_\nu}{k^4}\right)\equiv
D_{\mu\nu}.
\end{eqnarray}
This inverse $D_{\mu\nu}$ is transverse (i.e. it satisfies 
$k^\mu D_{\mu\nu}=0$) in the limit $\alpha\rightarrow 0$, even though
$\bar{\cal L}_{\mbox{gf}}$ is ill defined in this limit.
It is interesting to consider the possibility of
$D_{\mu\nu,\,\lambda\sigma}$ being transverse. From Eq. \eqref{e29}
it follows that
\begin{eqnarray}\label{eq36}
k^\mu  D_{\mu\nu,\,\lambda\sigma} &=& \frac{1}{k^2}\left[
(k_\lambda \eta_{\nu\sigma}+k_\sigma \eta_{\nu\lambda})({\bf C}^1+{\bf C}^3)
\right. \nonumber \\ &+& \left. 
k_\nu \eta_{\lambda\sigma}({\bf C}^2 + {\bf C}^4)
\right. \nonumber \\ &+& \left. 
\frac{k_\nu k_\sigma k_\lambda}{k^2} (2 {\bf C}^3 + {\bf C}^4 + {\bf C}^5)\right]. 
\end{eqnarray}
From Eqs.  \eqref{e30} we find that
\begin{subequations}\label{eq37}
\be
{\bf C}^1+ {\bf C}^3 = \frac{\beta^2}{4\xi}
\ee
\be
{\bf C}^2+ {\bf C}^4 = \frac{2}{d-2} \frac{\beta\gamma}{\alpha(\beta+\gamma)+\gamma(\alpha+\beta)}
\ee
\begin{eqnarray}
2 {\bf C}^3 + {\bf C}^4 + {\bf C}^5 &=& -\frac{\beta^2}{2\xi} + 
\frac{\beta\gamma}{[{\alpha(\beta+\gamma)+\gamma(\alpha+\beta)}]^2}
\nonumber \\  & & \!\!\!\!\!\!\!\!\!\!\!\!\!\!\!\!\!\!\!\!\!\!\!\!\!\!\!\!\!\!\!\!\!\!  \times  
\left(\frac{\alpha^2\beta \gamma }{\xi+\zeta}
-\frac{2[\alpha(\beta+\gamma)+\gamma(\alpha+ d \beta)]}{d-2} \right);
\end{eqnarray}
\end{subequations}
these all vanish if $\beta=0$ for all values of $\alpha$, $\gamma$,
$\xi$ and $\zeta$. If $\beta=0$, then
\be\label{eq50}
\eta^{\mu\nu} \left. D_{\mu\nu,\,\lambda\sigma}(k)\right|_{\beta=0} = 
-\frac{2}{d-2}\frac{1}{k^2}\left(\eta^{\lambda\sigma} -
  \frac{k^\lambda k^\sigma}{k^2}\right) 
\ee
showing that $D_{\mu\nu,\,\lambda\sigma}$ cannot be simultaneously
traceless and transverse with ${\cal L}_{\mbox{gf}}$ given by \eqref{e20},
irrespective of the values of $\alpha$, $\gamma$, $\xi$ and $\zeta$.

In general, from Eq. \eqref{e29} it follows that
\begin{eqnarray}\label{e51}
\eta^{\mu\nu} D_{\mu\nu,\,\lambda\sigma}(k)&=&
\frac{1}{k^2}\left[(2 {\bf C}^1 + d {\bf C}^2 +{\bf C}^4)\eta_{\lambda\sigma}
\right. \nonumber \\ &+& \left.
(4 {\bf C}^3 + d {\bf C}^4 + {\bf C}^5) \frac{k_\lambda k_\sigma}{k^2}\right]; 
\end{eqnarray}
from Eqs. \eqref{e30} if follows that
\begin{subequations}\label{e52}
\begin{eqnarray}
2 {\bf C}^1 + d {\bf C}^2 + {\bf C}^4 = -\frac{1}{k^2}\frac{2}{d-2}
\frac{\alpha(\beta+2\gamma)}{\alpha(\beta+\gamma)+\gamma(\alpha+\beta)}
\nonumber \\
\end{eqnarray}
\begin{eqnarray}
4 {\bf C}^3 + d {\bf C}^4 + {\bf C}^5 = 
\frac{1}{k^2} \frac{1}{[\alpha(\beta+\gamma)+\gamma(\alpha+\beta)]^2}
\nonumber \\ \times
\left[\frac{(\alpha \beta \gamma)^2}{\xi+\zeta}+
\frac{2\alpha }{d-2}(\beta+2\gamma)(\alpha(\beta+\gamma)+\gamma(\alpha+d\beta))
\right].
\nonumber \\
\end{eqnarray}
\end{subequations}
Thus if $\alpha=0$, we find that $D_{\mu\nu,\,\lambda\sigma}$
satisfies the traceless condition of Eq. \eqref{e6} for all values of
$\beta$, $\gamma$, $\xi$ and $\zeta$; if $\alpha=0$ then
\begin{eqnarray}\label{e54}
\left . D_{\mu\nu,\,\lambda\sigma}(k)\right|_{\alpha=0} &=&\frac{1}{k^2}\left(
P_{\mu\lambda}P_{\nu\sigma}+P_{\mu\sigma}P_{\nu\lambda}\right)
\nonumber \\
&+&\frac{\beta^2}{4k^4\xi}\Bigl[k_\mu k_\lambda \eta_{\nu\sigma}+k_\mu k_\sigma \eta_{\nu\lambda}
\Bigr.\nonumber \\ &+ &  \Bigl.                          
                              k_\nu k_\lambda \eta_{\mu\sigma}+k_\nu k_\sigma \eta_{\mu\lambda}
-\frac{4}{k^2}k_\mu k_\nu k_\lambda k_\sigma \Bigr]
\nonumber \\
&-&\frac{2}{(d-2) k^2}\left(\eta_{\mu\nu} - 2 \frac{k_\mu k_\nu}{k^2}\right)
\nonumber \\ & \times &
                      \left(\eta_{\lambda\sigma} - 2 \frac{k_\lambda k_\sigma}{k^2}\right)
\nonumber \\ 
&-&\frac{2}{k^6} k_\mu k_\nu k_\lambda k_\sigma, 
\nonumber \\ 
\end{eqnarray}
where 
\be\label{e55}
P_{\mu\nu} \equiv \eta_{\mu\nu} - \frac{k_\mu k_\nu}{k^2}. 
\ee
We note that in Eqs. \eqref{e30} the limits $\alpha\rightarrow 0$ and
$\beta\rightarrow 0$ do not commute and we have found that the former
limit leads to a traceless propagator that is not transverse while the
latter limit leads to a transverse propagator that is not traceless.

The DeDonder propagator \cite{tHooft:2002xp,Capper:1973pv}
\be\label{e56}
 D_{\mu\nu\lambda\sigma}(k)=
\frac{1}{k^2}\left[\eta_{\mu\lambda}\eta_{\nu\sigma}+\eta_{\nu\lambda}\eta_{\mu\sigma}
-\frac{2}{d-2}\eta_{\mu\nu}\eta_{\lambda\sigma}\right]
\ee
is recovered if  $\xi=1, \;\zeta=0, \;\alpha=\beta=-4\gamma=2$, 
or $\xi=1, \;\zeta=0, \;\alpha=-\beta=-2, \;\gamma=\infty$.

\section{Non-quadratic gauge fixing and the transverse-traceless gauge}
We start by observing that with
\be\label{e57}
{\cal L}_{\mbox{gf}} =-\frac{1}{\rho}\left(\partial_\mu h^{\mu\nu}\right)
\left(\partial^\nu h_{\nu\lambda}-\partial_\lambda h_{\nu}^{\nu}\right)
\ee
then the sum of the classical and gauge fixing Lagrangian contains the operator
\begin{eqnarray}\label{e58}
W_{\mu\nu,\,\lambda\sigma} = B_{\mu\lambda}B_{\nu\sigma}+B_{\mu\sigma}B_{\nu\lambda}
-2 B_{\mu\nu}B_{\lambda\sigma}\,;
\nonumber \\
B_ {\mu\nu} = \eta_{\mu\nu} k^2 - \left(1-\frac{1}{\rho}\right) k_\mu k_\nu
\end{eqnarray}
so that if 
$W_{\mu\nu,\,\alpha\beta} D^{(\rho)\alpha\beta}_{\;\;\;\;\;\; ,\lambda\sigma}=
\bar\Delta_{\mu\nu,\,\lambda\sigma}$ then
\begin{eqnarray}\label{e59}
D^{(\rho)}_{\mu\nu,\,\lambda\sigma}&=&
\frac{1}{2 k^2}\left(
P^\rho_{\mu\lambda}P^\rho_{\nu\sigma}+P^\rho_{\nu\lambda}P^\rho_{\mu\sigma}
\right. \nonumber \\ &-& \left.
\frac{2}{d-1}P^\rho_{\mu\nu}P^\rho_{\lambda\sigma}
\right)
\end{eqnarray}
where $P^\rho_{\mu\nu} = \eta_{\mu\nu} - (1-\rho) {k_\mu k_\nu}/{k^2}$. 
As $\rho\rightarrow 0$, from Eq. \eqref{e59} it follows that
\be\label{e60}
k^\mu D_{\mu\nu\; \lambda\sigma}^{(\rho=0)}=\eta^{\mu\nu} D_{\mu\nu\; \lambda\sigma}^{(\rho=0)}= 0.
\ee
However, Eq. \eqref{e57} is not of the form of Eq. \eqref{e20} and
hence the Faddeev-Popov procedure must be modified to accommodate such a
non-quadratic gauge fixing Lagrangian, which are needed if a
transverse-traceless propagator is to arise.

We begin by inserting two factors of ``1'' into Eq. \eqref{e8}; these
are
\begin{subequations}\label{e61t}
\be\label{e61}
1 = \int {\rm d}\vec\theta_1 
\delta(\undertilde{F}(\vec h + \alpha \undertilde{A} \vec\theta_1) - \vec p) 
\det(\alpha \undertilde{F} \undertilde A^{(0)})  
\ee
\be\label{e62}
1 = \int {\rm d}\vec\theta_2 
\delta(\undertilde{G}(\vec h + \alpha \undertilde{A} \vec\theta_2) - \vec q) 
\det(\alpha \undertilde{G} \undertilde A^{(0)})  
\ee
\end{subequations}
as well as another  ``1'' of the form
\be\label{e63}
1 = \pi^{-n}\int {\rm d}\vec p\,{\rm d}\vec q\,
{\rm e}^{-\frac{1}{\alpha}\vec p^{T} \undertilde{N} \vec q}{{\det}(\undertilde{N}/\alpha)}.
\ee
This leads to
\begin{eqnarray}\label{e64a}
Z & = &\pi^{-n}
\int{\rm d}\vec\theta_1 {\rm d}\vec\theta_2 \int{\rm d}\vec h 
\det(\alpha\undertilde{F}\undertilde{A}^{(0)})
\det(\alpha\undertilde{G}\undertilde{A}^{(0)}) 
\nonumber \\ &\times &
\det\left(\frac{\undertilde{N}}{\alpha}\right)
\exp\left\{-\vec h^T \undertilde{M}\vec h -\frac{1}{\alpha}
\left[\undertilde{F}(\vec h+\alpha \undertilde{A}^{(0)}\vec\theta_1)\right]^{T}
\right. \nonumber \\ & & \left.
\undertilde{N}
\left[\undertilde{G}(\vec h+\alpha \undertilde{A}^{(0)}\vec\theta_2)\right]\right\}.
\end{eqnarray}
We now make the shift $\vec h\rightarrow \vec h -\alpha\undertilde{A}^{(0)}\vec\theta_1$
in Eq. \eqref{e64a} and let $\vec\theta=\vec\theta_2-\vec\theta_1$ so
that by Eq. \eqref{e9}
\begin{eqnarray}\label{e64}
Z & = &\left(\frac{\alpha}{\pi}\right)^{n}
\int{\rm d}\vec\theta_1 \int{\rm d}\vec\theta \int{\rm d}\vec h 
\det(\undertilde{F}\undertilde{A}^{(0)})
\det(\undertilde{G}\undertilde{A}^{(0)}) 
\nonumber \\ &\times & 
\det(\undertilde{N})\exp\left\{-\vec h^T \left(\undertilde{M} + \frac{1}{\alpha}
\undertilde{F}^{T}\undertilde{N}\undertilde{G}\right)\vec h
\right. \nonumber \\ & - & \left.
\vec h^T \undertilde{F}^{T}\undertilde{N}\undertilde{G}\undertilde{A}^{(0)}\vec\theta
\right\}.
\end{eqnarray}
Dropping the infinite normalization factors in Eq. \eqref{e60} and
making the shift
\be\label{e65}
\vec h \rightarrow \vec h -\frac{1}{2}
\left(\undertilde{M} + \frac{1}{\alpha}
     \undertilde{F}^{T}\undertilde{N}\undertilde{G}\right)^{-1}
\left(\undertilde{F}^{T}\undertilde{N}\undertilde{G}\undertilde{A}^{(0)}\right)
\vec\theta
\ee
to diagonalize the exponential in Eq. \eqref{e64} in $\vec h$ and
$\vec\theta$, we obtain
\begin{eqnarray}\label{e66}
Z & = & 
\int{\rm d}\vec\theta \int{\rm d}\vec h 
\det(\undertilde{F}\undertilde{A}^{(0)})
\det(\undertilde{G}\undertilde{A}^{(0)}) 
\det(\undertilde{N})
\nonumber \\
&\times & 
 \exp\left\{-\vec h^T \left(\undertilde{M} + \frac{1}{\alpha}
\undertilde{F}^{T}\undertilde{N}\undertilde{G}\right)\vec h
\right. \nonumber \\ &+& \left.
\frac{1}{4}
\vec\theta^T
\left({\undertilde{A}^{(0)}}^{T}\undertilde{G}^{T}\undertilde{N}^{T}\undertilde{F}\right)
\left(\undertilde{M} + \frac{1}{\alpha}
     \undertilde{F}^{T}\undertilde{N}\undertilde{G}\right)^{-1}
\right. \nonumber \\ &\times& \left.
\left(\undertilde{F}^{T}\undertilde{N}\undertilde{G}\undertilde{A}^{(0)}\right)
\vec\theta
\right\}.
\nonumber \\
\end{eqnarray}
(We are assuming that $\undertilde{F}$, $\undertilde{G}$, $\undertilde{N}$ and
$\undertilde{A}^{(0)}$ are all independent of $\vec h$ so that no
Jacobian arises as a result of the change of variable in Eq. \eqref{e65}.)

If now we take
\be\label{e67}
\vec h^{T}\undertilde{F}^{T}\undertilde{N}\undertilde{G}\vec h =
h^{\mu\nu} F^T_{\mu\nu,\,\alpha} N^{\alpha\beta} 
           G_{\beta,\,\lambda\sigma}h^{\lambda\sigma}
\ee
with
\begin{subequations}\label{e68t}
\be\label{e68}
F^T_{\mu\nu,\,\alpha} = g_1 \eta_{\mu\nu} \partial_\alpha +
                            \eta_{\mu\alpha} \partial_\nu 
\ee
\be\label{e69}
G_{\beta,\,\lambda\sigma} = g_2 \eta_{\lambda\sigma} \partial_\beta +
                            \eta_{\lambda\beta} \partial_\sigma 
\ee
\be\label{e70}
N^{\alpha\beta} = \eta^{\alpha\beta}
\ee
\end{subequations}
then inverting the quadratic form 
$\undertilde{M} + \frac{1}{\alpha}
 \undertilde{F}^{T}\undertilde{N}\undertilde{G}$ 
leaves us with the coefficients in Eq. \eqref{e29} being
\begin{widetext}
\begin{subequations}\label{e71}
\be\label{genp1}
{\bf C}^1=1
\ee
\be\label{genp2}
{\bf C}^2=-2\frac{(g_2-g_1)^2+2(g_1+1)(g_2+1)\alpha}
{(d-1)(g_2-g_1)^2+2(d-2)(g_1+1)(g_2+1)\alpha}
\ee
\be\label{genp3}
{\bf C}^3={\alpha-1}
\ee
\be\label{genp4}
{\bf C}^4={2}
\frac{(g_2-g_1)^2+\left[4(g_1+1)(g_2+1)-g_1-g_2-2\right]\alpha}
{(d-1)(g_2-g_1)^2+2(d-2)(g_1+1)(g_2+1)\alpha}
\ee
\begin{eqnarray}\label{genp5}
{\bf C}^5 & = &
{\left[(d-1)(g_2-g_1)^2+2(d-2)(g_1+1)(g_2+1)\alpha\right]^{-1}}
\nonumber \\ &\times&
\left\{4\alpha\left[(g_1+g_2)(d-4)+(2 g_1 g_2+1)(d-3)-\left(g_1^2+g_2^2\right)(d-1)\right]
\right. \nonumber \\ &+&\left.
2(d-2)\left[(g_1-g_2)^2-\alpha ^2 (4(g_1+1)(g_2+1)-1) \right]\right\}
\end{eqnarray}
\end{subequations}
\end{widetext}
From these expressions we see that the limits $g_2\rightarrow g_1$ and
$\alpha\rightarrow 0$ do not commute. If we take the limit
$\alpha\rightarrow 0$, with $g_2\neq g_1$, the propagator becomes
independent of $g_1$ and $g_2$, and we obtain the transverse and
traceless propagator. On the other hand, if we set $g_2=g_1$, the
resulting propagator is not transverse and traceless even for
$\alpha=0$. This is another verification of the impossibility of
obtaining the transverse and traceless propagator using the quadratic
gauge fixing where $g_1=g_2$.
This general gauge fixing also can be used to find the DeDonder
propagator of Eq. \eqref{e56} by taking $g_1=g_2=-1/2$ and $\alpha=1$.
It is also interesting to note that for $d=2$, and arbitrary values of
$g_1$, $g_2$ and $\alpha$, Eqs. \eqref{e71} are well
defined while the DeDonder propagator of Eq. \eqref{e56} is not.

The determinants in Eq. \eqref{e66} can all be exponentiated using
Grassmann quantities $\vec c$, $\vec{\bar c}$, $\vec b$, $\vec{\bar b}$,
$\vec k$ and $\vec{\bar k}$, so that
\begin{eqnarray}\label{e76}
Z & = & 
\int{\rm d}\vec\theta \int{\rm d}\vec h
\int{\rm d}\vec c\,{\rm d}\vec{\bar c}
\int{\rm d}\vec b\,{\rm d}\vec{\bar b}
\int{\rm d}\vec k\,{\rm d}\vec{\bar k}
\nonumber \\ &\times & 
\exp\left\{-\vec h^T \left(\undertilde{M} + \frac{1}{\alpha}
\undertilde{F}^{T}\undertilde{N}\undertilde{G}\right)\vec h
\right. \nonumber \\ &-& \left.
\vec{\bar b}\left(\undertilde{F}\undertilde{A}^{(0)}\right)\vec b
-\vec{\bar c}\left(\undertilde{G}\undertilde{A}^{(0)}\right)\vec c
-\vec{\bar k}\undertilde{N}\vec k 
\right.\nonumber \\ & + & \left.
\frac{1}{4}
\vec\theta^T
\left({\undertilde{A}^{(0)}}^{T}\undertilde{G}^{T}\undertilde{N}^{T}\undertilde{F}\right)
\left(\undertilde{M} + \frac{1}{\alpha}
     \undertilde{F}^{T}\undertilde{N}\undertilde{G}\right)^{-1}
\right.\nonumber \\ &\times & \left.
\left(\undertilde{F}^{T}\undertilde{N}\undertilde{G}\undertilde{A}^{(0)}\right)
\vec\theta
\right\}
\end{eqnarray}
up to a normalization factor.

The gauge fixing of Eq. \eqref{e57} corresponds to $g_2=-1$,
$g_1\rightarrow 0$ and $\alpha=\rho$. In this case, the determinant 
$\det(\undertilde{G}\undertilde{A}^{(0)})=0$ and hence the ghost
Lagrangian $\vec{\bar b}(\undertilde{G}\undertilde{A}^{(0)})\vec b$ itself
possess a gauge invariance 
$\vec b\rightarrow\vec b +  \undertilde{B}^{(0)} \vec\omega$, where
$\vec\omega$ is a Grassmann gauge function. Following the
Faddeev-Popov procedure, we find that
\begin{eqnarray}
\int{\rm d}\vec b\,{\rm d}\vec{\bar b} 
\exp\left[-\vec{\bar b}\left(\undertilde{G}\undertilde{A}^{(0)}\right)\vec b \right]
 =   \int{\rm d}\vec b\,{\rm d}\vec{\bar b} \int{\rm d}\vec\beta\,{\rm d}\vec{\bar\beta}
 \int{\rm d}\vec{\cal H}
\nonumber \\ 
\times\exp\left[-\vec{\bar b}\left(\undertilde{G}\undertilde{A}^{(0)}
+\undertilde{\Gamma}^T\undertilde{\eta}\undertilde{\Gamma}\right)\vec b 
\right. \nonumber \\ - \left.
\vec{\undertilde{\beta}}^T\left(\undertilde{\Gamma}\undertilde{B}^{(0)}\right)\vec\beta 
-\vec{\cal H}^T\undertilde{\eta}\vec{\cal H}\right],
\nonumber \\
\end{eqnarray}
where $\vec\beta$ and $\vec{\bar\beta}$ are complex Faddeev-Popov
ghosts and $\vec{\cal H}$ is a real Nielsen-Kallosh ghost (with
neither of these being Grassmann). We will not consider this gauge
fixing further in order to avoid having to introduce these 
``ghosts of ghosts''.

The field $\vec \theta$ appearing in Eq. \eqref{e76} is a non-trivial
propagating field that has no analogue in the usual Faddeev-Popov
procedure. The propagator for $\vec\theta$ with the gauge fixing
chosen to be Eqs. \eqref{e68t} and the gauge transformation given by
\eqref{e18} is
\begin{eqnarray}\label{e78}
D_\theta^{\mu\nu}(k) &=&
\frac{1}{\alpha}\frac{1}{k^4}\left\{\eta^{\mu\nu} 
- \left[\left(1-\frac{1}{4(g_1+1)(g_2+1)}\right) 
\right.\right. \nonumber \\ &-& \left.\left.
\frac{1}{8\alpha}\frac{d-1}{d-2}\left(\frac{1}{g_1+1}
                                     +\frac{1}{g_2+1}\right)^2
\right] \frac{k^\mu k^\nu}{k^2} \right\}.
\nonumber \\ & & 
\end{eqnarray}

Upon performing the functional integrals over the fields 
$\vec\theta$, $\vec h$, $\vec{\bar c}$, $\vec c$, and $\vec{\bar b}$, $\vec b$
(taking $\undertilde{N}=1$) in \eqref{e76} we find that
\begin{eqnarray}\label{e79}
Z & = & 
{\det}^{-1/2}\left(\undertilde{M}+\frac{1}{\alpha}\undertilde{F}^T\undertilde{G}\right)
\det{\left(\undertilde{F}\undertilde{A}^{(0)}\right)}
\det{\left(\undertilde{G}\undertilde{A}^{(0)}\right)}
\nonumber \\ &\times &
{\det}^{-1/2}\left[\left({\undertilde{A}^{(0)}}^T\undertilde{G}^T\undertilde{F}\right)
\left(\undertilde{M}+\frac{1}{\alpha}\undertilde{F}^T\undertilde{G}\right)^{-1}
\right. \nonumber \\ & \times & \left.
\left(\undertilde{F}^T\undertilde{G}\undertilde{A}^{(0)}\right)\right].
\end{eqnarray}
With Eqs. \eqref{e18} and  \eqref{e68t} these determinants become
\begin{eqnarray}\label{e80}
Z & = & 
\left[\frac{3(g_1-g_2)^2+4\alpha(g_1+1)(g_2+1)}{\alpha^5}\,{(\det\partial^2)^{10}}\right]^{-\frac{1}{2}}
\nonumber \\ &\times &
\left[2 (g_1+1) (\det\partial^2)^4\right]\left[2 (g_2+1) (\det\partial^2)^4\right]
\nonumber \\ &\times &
\left[16\frac{\alpha^5(g_1+1)^2(g_2+1)^2}{3(g_1-g_2)^2+4\alpha(g_1+1)(g_2+1)}
\,{(\det\partial^2)^{8}}\right]^{-\frac{1}{2}} 
\nonumber \\ & &
\end{eqnarray}
which reduces to
\be\label{e81}
Z=(\det\partial^2)^{-1}.
\ee
This indicates that there are in fact just two Bosonic degrees of
freedom, as the contribution of a single scalar degree of freedom is
\be
\int{\rm d}\phi {\rm e}^{\phi\partial^2\phi} = (\det\partial^2)^{-1/2}
\ee
These two degrees of freedom are of course the transverse
polarizations of the free graviton. The free energy is thus given by
\cite{lebellac:book96}
\begin{eqnarray}
&\displaystyle{-T V \log{Z^{(0)}} = {\Omega(T)}} 
\nonumber \\ &
\displaystyle{= 2\,V \int \frac{{\rm d}^3 k}{(2\pi)^3} 
\left[\frac{|\vec k|}{2} + T\, \log\left(1-{\rm e}^{-(k/T)}\right)\right]},
\end{eqnarray}
the factor of two coming from the two degrees of freedom.

We first note that in Eq. \eqref{e81} all dependence on the gauge
parameters has vanished. We also see that from Eq. \eqref{e80} all
determinants in Eq. \eqref{e79} are non-zero.

We now consider the situation in which the spin-two field is no longer
a free-field due to the self-interactions. The path integral to be
considered then is not in the form of Eq. \eqref{e8}; we now must
examine
\be\label{e84}
Z_I = \int{\rm d}\vec h\exp\left[-\vec h^T\undertilde{M}\vec h - S_I(\vec h)\right]
\ee
where $S_I(\vec h)$ is at least cubic in $\vec h$. The argument of the
exponential in Eq. \eqref{e84} is now invariant under a transformation
\be\label{e85}
\vec h \rightarrow (\vec h)_{\vec\omega} = 
\vec h + \alpha\undertilde{A}(\vec h)\vec\omega+{\cal O}(\vec\omega^2),
\ee
where $\vec\omega$ is arbitrary and $\undertilde{A}(\vec h)$ now
depends on $\vec h$, with 
$\undertilde{A}(\vec h) = \undertilde{A}^{(0)}+{\cal O}(\vec h)$. 
Factors of ``1'' are now inserted into Eq. \eqref{e84}, using
Eqs. \eqref{e61t} with $(\vec h)_{\vec\omega}$ replacing 
$\vec h + \alpha\undertilde{A}^{(0)}\vec\theta$, and keeping
Eq. \eqref{e63}. Thus in place of Eq. \eqref{e64} we obtain (up to a
normalization factor)
\begin{eqnarray}\label{e86}
Z_I & = & \int{\rm d}\vec\theta\int{\rm d}\vec h
\det(\undertilde{F} \undertilde{A}(\vec h)) 
\det(\undertilde{G} \undertilde{A}(\vec h)) \det(\undertilde{N})
\nonumber \\ & & 
\exp\left[-\vec h^T\undertilde{M}\vec h - S_I(\vec h)
-\frac{1}{\alpha}\vec h^T\undertilde{F}^T\undertilde{N}\undertilde{G}
(\vec h)_{\theta}\right],
\nonumber \\ & & 
\end{eqnarray}
where 
$(\vec h)_{\theta} = \left((\vec
  h)_{\theta_2}\right)_{\theta_1^{-1}}\approx\vec h+
\alpha\undertilde{A}^{(0)}(\vec\theta_2-\vec\theta_1)$. The shift of
Eq. \eqref{e65} can again be used to diagonalize the terms appearing
in the argument of the exponential in \eqref{e86} that are quadratic
in $\vec h$ and $\vec\theta$, but this shift also induces extra
vertices involving the field $\vec\theta$, as Eq. \eqref{e65} is not
of the form of a gauge transformation. However, as has been noted
above, the gauge fixing of Eqs. \eqref{e68t} results in 
$(\undertilde{M}+\frac{1}{\alpha}\undertilde{F}^T\undertilde{N}\undertilde{G})^{-1}$
being traceless and transverse as $\alpha\rightarrow 0$, so that the
shift
$-\frac{1}{2}(\undertilde{M}+\frac{1}{\alpha}\undertilde{F}^T\undertilde{N}\undertilde{G})^{-1}
(\undertilde{F}^T\undertilde{N}\undertilde{G}\undertilde{A}^{(0)})\vec\theta$
appearing in Eq. \eqref{e68} is formally of order $\alpha$. Keeping in
mind that the propagator for the field $\vec\theta$ in Eq. \eqref{e78}
has contributions of order $1/\alpha$ and $1/\alpha^2$ (though the
latter disappears if $1/(g_1+1)+1/(g_2+1)=0$), we see that as
$\alpha\rightarrow 0$ the contribution of these extra vertices is
reduced.

\section{Discussion}
We have examined the most general covariant quadratic gauge fixing
Lagrangians for a spin-two gauge field and have shown that none of them
can be used to obtain the transverse-traceless propagator for this
field. Non-quadratic gauge fixing Lagrangians can however be used to
obtain this propagator, and we have shown that their systematic
introduction results in an unconventional ghost contribution to the
effective action. In a different context Drummond and Shore 
have also considered non-quadratic gauge fixing Lagrangians \cite{Drummond:1977uy,Shore:1977df}.

It would be worth to derive the 
WTST \cite{Ward:1950xp,Takahashi:1957xn,Slavnov:1972fg,Taylor:1971ff,Capper:1974vb}
and BRST \cite{Becchi:1976nq} identities when these non-quadratic
gauge fixing Lagrangians are used and to verify them by explicit
calculation of loop diagrams.
As a first step towards the calculation of more involved perturbative
quantities, one may consider the one-loop contributions to the thermal
energy-momentum tensor. Since this result is known in
the usual formulation of thermal gravity \cite{Gribosky:1988yk,Rebhan:1990yr}, one can
verify the consistence of the non-quadratic gauge fixing approach
in a specific scenario such that the interactions cannot be neglected.

\begin{figure}[h!]
\begin{center}
$
\begin{array}{c}
\includegraphics[scale=0.5]{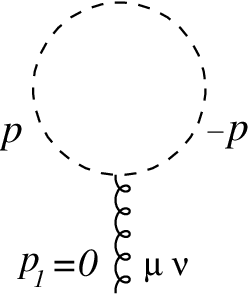}\\ \\ (a)
\end{array}
$
$
\begin{array}{c}
\includegraphics[scale=0.5]{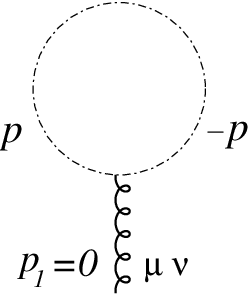}\\ \\ (b)
\end{array}
$
\\
$
\begin{array}{c}
\includegraphics[scale=0.5]{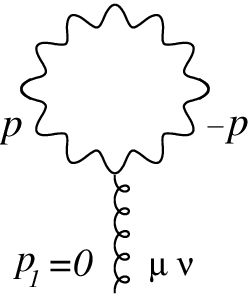}\\ \\ (c)
\end{array}
$
$
\begin{array}{c}
\includegraphics[scale=0.5]{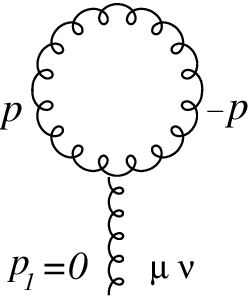}\\ \\ (d)
\end{array}
$
\end{center}
\caption{Diagrams which contribute to the thermal one-graviton function.
The dashed and dashed-doted lines represent the ghost fields $b$ and $c$,
respectively. The wavy lines represent the $\theta$ field and the
curly lines represent gravitons.}\label{figtad1}
\end{figure}

The general relation between the one-graviton function $\Gamma^{\mu\nu}$
and the energy-momentum tensor $T^{\mu\nu}$ is such that
\be
\Gamma^{\mu\nu}=\frac{\delta \Gamma}{\delta h_{\mu\nu}} = -\frac{1}{2}\sqrt{-g}\,  T^{\mu\nu}  ,
\ee
where $\Gamma$ is the one-loop thermal effective action. 
In the figure \ref{figtad1} we shown the lowest order diagrams which contribute to $\Gamma^{\mu\nu}$. 
In order to compute these diagrams we need the propagators,
derived in the previous sections, as well as the interactions vertices.
Let us first consider  the diagram (a) in figure \eqref{figtad1}. 
The ghost-graviton vertices can be obtained from the
gauge-invariant completion of the quantity $-\vec{\bar b}(\undertilde{F}\undertilde{A}^{(0)})\vec b$
(see Eq. \eqref{e76}) so that $\undertilde{A}^{(0)}$ is replaced by  
\begin{eqnarray}\label{gh2}
{A}_{\mu\nu\rho} &=& g_{\mu\rho}\partial_\nu+g_{\nu\rho}\partial_\mu+(\partial_\rho g_{\mu\nu})
\nonumber \\
&=&\eta_{\mu\rho}\partial_\nu+\eta_{\nu\rho}\partial_\mu 
\nonumber \\
&+& h_{\mu\rho}\partial_\nu+h_{\nu\rho}\partial_\mu + (\partial_\rho h_{\mu\nu}). 
\end{eqnarray}
In this way, both the propagator and the interaction vertex can be read from the
the Lagrangian density
\be\label{gh0}
-\bar b_\lambda(F A)^\lambda_\rho b^\rho.
\ee
Using Eq. \eqref{e68}, we obtain
\begin{eqnarray}\label{hb0}
\bar b_\lambda\, (\undertilde F\undertilde A)^\lambda_\rho\,  b^\rho &=&
\bar b_\lambda\left[(2g_1+1)\partial^\lambda\partial_\rho+\delta^\lambda_\rho\partial^2 \right] b^\rho 
\nonumber \\
&+& \bar b_\lambda\left[g_1 (2 \partial^\lambda  h^\nu_\rho \partial_\nu
                           + \partial^\lambda (\partial_\rho h^\nu_\nu)) 
\right. \nonumber \\ &+& \left.  
\partial^\nu h^\lambda_\rho \partial_\nu+\partial_\nu h^\nu_\rho \partial^\lambda
+ \partial_\nu  (\partial_\rho h^{\nu\lambda})   \right] b^\rho. 
\nonumber \\ & &
\end{eqnarray}
Let us now associate momenta $p_1$,  $p_2$ and $p_3$,  in momentum space, 
respectively to the graviton field $h_{\mu\nu}$,  to $\bar b_\lambda$ and to 
$ b_\rho$.  This yields the following interaction vertex for the $b$ ghost
\begin{eqnarray}\label{hb1}
V^{\mu\nu;\, \lambda\rho}_{ b}(p_1, p_2, p_3) & = &
\frac{g_1}{2}\left(2 p_2^\lambda
  p_3^\mu\eta^{\nu\rho}+p_2^\lambda p_1^\rho \eta^{\mu\nu}\right) 
\nonumber \\ &+& 
\frac{1}{2}\left(
p_2\cdot p_3 \eta^{\mu\lambda}\eta^{\nu\rho} + p_2^\mu p_3^\lambda\eta^{\nu\rho}
\right.\nonumber \\ &+& \left. 
p_2^\mu p_1^\rho\eta^{\nu\lambda}
\right) + (\mu\leftrightarrow\nu). 
\end{eqnarray}
In the diagrams of figure \ref{figtad1} the momenta are such that
$p_1=0$ and $p_2=-p_3=p$ so that
\begin{eqnarray}
V^{\mu\nu;\, \lambda\rho}_{ b}(0, p, -p) & = &
-\frac{\eta^{\nu\rho}}{2}\left[(2g_1+1)p^\mu p^\lambda+p^2\eta^{\mu\lambda}\right]
\nonumber \\ &-& 
\frac{\eta^{\mu\rho}}{2}\left[(2g_1+1)p^\nu p^\lambda+p^2\eta^{\nu\lambda}\right]. 
\end{eqnarray}
When we contract with the ghost propagator, which is given by the
inverse of the first term in Eq. \eqref{hb0}, we obtain
\begin{eqnarray}
V^{\mu\nu;\, \lambda\rho}_{ b}(0, p, -p) 
\left[(2g_1+1)p^\lambda p^\rho+\eta^{\lambda\rho} p^2 \right]^{-1}
= 
\nonumber \\
-\frac{1}{2}\left(\eta^{\nu\rho}\delta^\mu_\rho+
                    \eta^{\mu\rho}\delta^\nu_\rho \right)
= -\eta^{\mu\nu}. 
\end{eqnarray}
The same can be done with the diagram (b) in figure \eqref{figtad1},
which is associated with the ghost field $c$, so that
\be
V^{\mu\nu;\, \lambda\rho}_{c}(0, p, -p) 
\left[(2g_2+1)p^\lambda p^\rho+\eta^{\lambda\rho} p^2 \right]^{-1}
= 
\nonumber \\
-\eta^{\mu\nu}. 
\ee
We now have to integrate these expressions over ${\rm d}^{d-1} p$ and sum over
the Matsubara frequencies $p_0=2\pi n T$.  Then,  the dimensionally
regularized integral will yield a zero result for both ghosts loops. 

Let us now consider the contribution of the $\theta$ field. 
The vertex in the diagram (c) in figure \eqref{figtad1} is the sum
of two types of contributions.  The first contribution comes from the order $h$ terms 
when we replace $\tilde A^{(0)}$ in Eq. \eqref{e76} by  the Eq.  \eqref{gh2}.
This type of contribution will also yield an expression for the integrand which is proportional to
$\eta_{\mu\nu}$.  Indeed,  as in
the case of the ghost fields $b$ and $c$,  the only relevant part 
of the interaction vertex is the one which has a zero momentum external
graviton,  so that the order $h$ terms in $A$ will yield a contribution
proportional to the inverse of the propagator.  Therefore,  this part of
the interaction will not contribute to the energy-momentum tensor. 

The second part of the interaction between the $\theta$ and $h$ fields
arises when the cubic term in the interaction Lagrangian is modified
by the shift given by \eqref{e65}. In order to compute this contribution
we employ the known expression for the three graviton vertex
\cite{Brandt:1992dk} and contract two of its external legs 
with the operator on the right hand-side of \eqref{e65}.
Finally, contracting the resulting expression 
with the $\theta$ propagator in \eqref{e78} we have obtained
\be\label{theta100}
\frac{(d-3)(g_1-g_2)^2}{(d-1)(g_1-g_2)^2+2\mu(d-2)(g_1+1)(g_2+1)}
\frac{p^\mu p^\nu}{p^2}
\ee

The last diagram of  in figure  \eqref{figtad1} has the usual
interaction vertex contracted with the general propagator given by
Eqs. \eqref{e29} and \eqref{e71}. A straightforward calculation yields
(we have employed the symbolic computer package HIP \cite{Hsieh:1991ti})
\be\label{theta101}
\frac{(d-3)(d-2)[(d+1)(g_1-g_2)^2+2 \alpha d (g_1+1)(g_2+1)]}
{2[(d-1)(g_1-g_2)^2+2\mu(d-2)(g_1+1)(g_2+1)]}
\frac{p^\mu p^\nu}{p^2}
\ee
Adding the two previous expressions, we obtain
\be\label{theta102}
\frac{d(d-3)}{2}\frac{p^\mu p^\nu}{p^2}. 
\ee
All the gauge parameter dependence has been canceled in the
final expression for the integrand of the one point function and the
result agrees with the known result in the 
DeDonder gauge.  Of course this gauge independent result is expected
for a physical quantity like the energy-momentum tensor.
This rather simple calculation indicates that the interactions can be
taken into account consistently in the double gauge fixing formulation.
Since this calculation has been done without restricting the values of 
the gauge parameters $\alpha$, $g_1$ and $g_2$, it also holds in the
particular case of the TT graviton propagator. The combination of
the expressions \eqref{theta100} and \eqref{theta101} yielding the
gauge invariant result shows how the modes associated with the
$\theta$ and $h$ fields combines to produce the correct result.

Another interesting application of the TT gauge can be made to study
the thermal loop-corrections to the free-energy in quantum
gravity. This would allow for a simple and physical analysis of the
Jeans-like instabilities which develop at non-zero temperature.
Work on this topic is in progress.

\acknowledgements

F. Brandt and J. Frenkel would like to thank Fapesp and Cnpq for
financial support.
{\hbox{D. G. C. McKeon}} would like to thank Fapesp and the University of São
Paulo for its generous support and warm hospitality while most of this
was done, and Roger MacLeod for a helpful suggestion.


\end{document}